\documentclass[aip,jcp,preprint,floatfix]{revtex4-1}
\usepackage[T1]{fontenc}
\usepackage[utf8]{inputenc}
\usepackage{amsmath}
\usepackage{booktabs}
\usepackage{graphicx}
\usepackage{dcolumn}
\usepackage{multirow}
\usepackage{braket}
\usepackage{color}
\usepackage{nicefrac}
\usepackage{gensymb}
\usepackage{enumitem}
\usepackage[version = 3]{mhchem}
\usepackage{paralist}
\usepackage[para]{threeparttable}

\newcommand{\beq}{\begin{equation}}
\newcommand{\eeq}{\end{equation}}

\begin{document}
\title{First-principle interaction potentials for metastable He($^3$S) and Ne($^3$P) with closed-shell molecules. Application to Penning-Ionizing systems}

\author{Micha{\l} Hapka}
\author{Grzegorz Cha{\l}asi{\'n}ski}
\affiliation{Faculty of Chemistry, University of Warsaw, 02-093 Warsaw, Pasteura 1, Poland}
\author{Jacek K{\l}os}
\affiliation{Department of Chemistry and Biochemistry, University of Maryland, College Park, MD 20742-2021, USA}
\author{Piotr S. \.Zuchowski}
\email{pzuch@fizyka.umk.pl}
\affiliation{Institute of Physics, Nicolaus Copernicus University, Grudziądzka 5, 87-100 Toru\'n, Poland}

\begin{abstract}

We present new interaction potential curves, calculated from first-principle, for the He$(^3$S)$\cdots$H$_2$ and He$(^3$S)$\cdots$Ar systems, relevant in Penning's ionization experiments. Two different approaches were applied: supermolecular using coupled cluster theory (CC) and perturbational within symmetry-adapted perturbation theory (SAPT). Both methods gave consistent results and the potentials were used to determine the positions of shape resonances in low collision energy scattering regime. We found a good agreement with the most recent scattering experiment of Henson {\em et al.} [Science {\bf 338}, 234 (2012)]. 

In addition, we investigated two other dimers,  composed of metastable Ne and ground state He and Ar atoms. For the Ne$(^3$P)$\cdots$He system a good agreement between CC and SAPT approaches was obtained. The Ne$(^3$P)$\cdots$Ar dimer was described only with SAPT, as CC gave divergent results. Ne$^*$ systems exhibit extremely small electronic orbital angular momentum anisotropy of the potentials. We attribute this effect to the screening of the open 2$p$ shell by the singly occupied 3$s$ shell.  

\end{abstract}

\maketitle

\section{Introduction}

The quest for taking control over chemical reactions and scattering processes by manipulating the movement of molecules led in the last decade to a rapid developement of methods which use the electric~\cite{Bethlem:1999,Bethlem:ag:2002,Sommer:2010} or magnetic fields~\cite{Narevicius:2008,Narevicius:2012} control combined with molecular beam techniques.~\cite{Meerakker:2012}
The possibility for quantum control opens in the low-energy regime, when the kinetic energy of the molecule becomes comparable to the perturbations due to the external electromagnetic field.~\cite{Krems:2008} 
Under these conditions the scattering is dominated by several partial waves only, facilitating studies of such phenomena as scattering resonances.
Moreover, elaborate detection techniques allow to study the state-resolved cross sections and the propensity rules, governing the distribution
of the inelastic scattering products.~\cite{Kirste:2010}
Finally, the experiments on low-energy scattering  provide unique opportunities for verification of the potential energy surfaces of colliding species, which is particularly beneficial for developing theory of intermolecular forces.

Recent progress in this field involved theoretical works in the scattering theory~\cite{Jachymski:2013} and experiments employing metastable noble gas atoms interacting with other species.~\cite{Doret:2009,Henson:2012} Studies of Henson {\it et al.}~\cite{Henson:2012} revived the interest in Penning ionization reaction (PI) -- in their work a beam of metastable helium atoms in the $2 ^3$S state, with their velocity controlled by a time-varying magnetic field, has been merged with a supersonic beam seeded with H$_2$ molecules or Ar atoms. Thanks to the velocity control, for the first time it was possible to observe sub-Kelvin scattering resonances with sub-milikelvin resolution.

Understanding of PI is very important since this process occurs in plasma~\cite{Fridman:2008,Liu:2010} and surface chemistry.~\cite{Harada:97} The reaction takes place when molecules collide with species excited to a very high energies, such as metastable $2 ^3$S He or $^3$P Ne atoms. Provided that an atom is excited above the ionization threshold of the molecule, the collision quenches the atom down to the ground state releasing internal energy, which in turn can ionize the molecule: \ce{A^* + B -> A + B^+ + e^-}.

Metastable He and Ne atoms have also been of great interest to cold matter physics, in particular the Bose-Einstein condensate of metastable He atoms has been obtained.~\cite{DosSantos:2001,Doret:2009} It was demonstrated that light atoms at very low temperatures could be used for efficient sympathetic cooling of molecules~\cite{Wallis:2011,Zuchowski:2011} and metastable helium atoms in $\mu$K regime could be considered as coolants, provided that the anisotropy of the potential energy surface is not very large.~\cite{Zuchowski:2008,Soldan:2009}

Because of the lack of interest, the theoretical effort to improve and develop potentials for metastable atoms interacting with molecules has been abandoned for more than 10 years. However, new perspectives offered by the experiments with metastable atoms require knowledge of accurate potential energy surfaces of those systems.

There is a principal difficulty in treating electronic states of the A$^*$-B type that are submerged in the continuum of states of the (AB)$^{+}$+$e$ type: the variational principle, which is the fundamental tenet of quantum chemistry, is expected to drive such states down either to the ground state, or one of the excited states or  -- because of the coupling with continuum states -- to some delocalized state corresponding to the fragmentation into an ion, a molecule, and a free electron. Thus, during the wave function optimization one should seek for a constraint minimum of the energy functional that does not obey the {\em Aufbau} rule.

Interaction potentials for systems undergoing PI  have been obtained with either {\it ab initio} or semi-empirical methods. Fully {\it ab initio} approaches encompass the ``stabilization method'' (e.g., see Ref. \onlinecite{Miller:70}), the configuration interaction approach with embedded Feshbach projector operators~\cite{Hickman:77} and the Siegert-eigenvalues techniques.~\cite{Isaac:78} It is, nevertheless, the semi-empirical method based on fitting parameters of model potentials to the experimental results that enabled construction of the most accurate PI interaction potentials. Different models have been suitably adapted over the years: a one-electron model potential by Siska~\cite{siska1979one}, Lennard-Jones potentials and so called piecewise potentials, constructed with components insuring a proper short- and long-rage behaviour. Ultimately, in order to describe He$^*$(2~$^1$S)$\cdots$H$_2$ interaction Martin and Siska introduced a non-piecewise potential surface function based on a modified Tang-Toennies model~\cite{Martin:88} (for a thorough review of previous theoretical works see Ref.~\onlinecite{Siska:93}).

The most straightforward among many-electron {\it ab initio} methods is the ``stabilization method'' (SM). The wave function is expanded in an atomic/molecular basis set, followed by diagonalization of the Hamiltonian matrix, thus all standard procedures designed for electronic bound states may be applied. An obvious disadvantage of SM is a possibility of obtaining continuum-like solutions in close proximity to the true resonances. Fortunately, such spurious solutions may often be identified due to their strong basis-set dependence.

In this work we present state-of-the art calculations of intermolecular potentials for He*- and Ne*-molecule dimers.
To this end, we applied two independent methods: \begin{inparaenum}[(i)] \item the supermolecular approach based on the coupled cluster method with judiciously prepared reference states, and \item the perturbation approach involving symmetry-adapted perturbation theory (SAPT) for open-shell species.~\cite{Hapka:2012} \end{inparaenum}

Previous works exploring application of different formulations and levels of SAPT expansion to excited-state systems are sparse: they can be found in Refs.~\onlinecite{Hemert:79,Greg:1980,Korona:99,Przybytek:2004,Zuchowski:2003}. Interactions of a metastable helium atom were addressed in two of them -- in Ref.~\onlinecite{Przybytek:2004} Przybytek {\it et al.} performed a detailed study of the convergence behaviour of SAPT expansion for He($^3$S)$\cdots$H interaction, whereas {\.Z}uchowski {\it et al.}~\cite{Zuchowski:2003} calculated the dispersion energy contribution in the random-phase approximation for the He($^3$S) dimer.

In the case of PI reaction, the SAPT methodology is appealing, as it ensures a {\em fixed} number of electrons at each monomer. The issue of concern, however, is the convergence of SAPT, as shown e.g. for dimers of metals.~\cite{Patko:2007,Hapka:2012}

The systems which we investigate in this paper either undergo Penning ionization: He$^*\cdots$H$_2$, He$^*\cdots$Ar (studied also by Henson {\em et al.}~\cite{Henson:2012}), and Ne$^*\cdots$Ar~\cite{Gregor:81}, or provide an auxiliary model for testing SAPT performance for such systems, as the Ne$^*\cdots$He dimer.

\section{Computational details}
For both complexes of metastable helium we were able to perform supermolecular unrestricted coupled cluster singles and doubles with noniterative triples correction (UCCSD(T)) calculations.~\cite{Knowles:93} It should be stressed, that convergence of CCSD(T) for excited states calculations cannot be taken for granted. Therefore, we performed them carefully by the following procedure.
First, we ran separate monomer calculations where desired states of monomers were set: in our case, one monomer was a metastable helium, the other was either the ground state H$_2$ molecule or an Ar atom. The resulting orbitals were then used to build the starting wave function of
the dimer as a product of two monomer functions. During the optimization of the ROHF wave function we did not use of the {\em Aufbau} 
principle, and in each iteration of ROHF, orbitals were reordered to obtain maximum overlap with closed-shell/open-shell/virtual spaces from the previous iteration.
For the Ne$^*\cdots$He system the UCCSD(T) calculations were straightforward since the state is the first triplet state. The UCCSD(T) results subsequently served as a benchmark for the performance of SAPT calculations. It should be noted, that the CC approach did fail for the potential curves of the Ne$^*\cdots$Ar system.

Both metastable He($^3$S) and Ne($^3$P) are high-spin open-shell atoms and can be treated with SAPT based on restricted open-shell or unrestricted Hartree-Fock and Kohn-Sham description of the monomers, SAPT(ROHF) and SAPT(ROKS)~\cite{Zuchowski:2008a} or SAPT(UHF) and SAPT(UKS)~\cite{Hapka:2012}, respectively. The method is based on the expansion of the interaction energy in the power series of the intermolecular interaction operator, which up to the second order takes the form:
\begin{equation}
\begin{split}
E_{\rm int}^{\rm SAPT} & = E_{\rm elst}^{(1)} + E_{\rm exch}^{(1)} + E^{(2)}_{\rm ind}+ E_{\rm disp}^{(2)} + E^{(2)}_{\rm exch-ind}  + E_{\rm exch-disp}^{(2)},
\end{split}
\label{tot:ks}
\end{equation}
where on the right-hand side the consecutive terms represent the electrostatic, first-order exchange energy, induction and dispersion energies, and 
second-order exchange-induction and exchange-dispersion energies. In this work we choose the unrestricted, SAPT(UHF) and SAPT(UKS) formalisms. The second-order terms have been calculated at the coupled level of theory.~\cite{Hapka:2012} In SAPT(UKS) calculations the PBE0 xc functional \cite{Perdew:96, Adamo:99} was asymptotically corrected~\cite{Gruning:2001}, with vertical ionization potentials calculated using a difference method.

For UCCSD(T) calculations double augmented, correlation consistent quadrupole-zeta basis set (d-aug-cc-pVQZ) of Dunning {\it et al.}~\cite{dunning1989gaussian} was chosen for helium and hydrogen atoms. For Ar we used aug-cc-pVQZ basis set. Ne($^3$P)$\cdots$He complex was described with d-aug-cc-pVQZ basis supplemented with 3$s$3$p$2$d$2$f$ bond functions (with exponents equal to: $sp$ 0.9, 0.3, 0.1, $df$ 0.6, 0.2) placed at the midpoint of the intermolecular distance. In SAPT calculations different basis sets had to be chosen in order to insure convergence of the metastable monomers. For a metastable helium, d3Z basis set of Przybytek~\cite{Przybytek:2008} was selected. Hydrogen was described by the aug-cc-pVQZ basis. In the case of He($^3$S)$\cdots$Ar complex we chose the aug-cc-pVQZ basis set for argon and a set of 3$s$3$p$2$d$ bond functions was added in order to saturate the dispersion interaction. In Ne($^3$P)$\cdots$Ar aug-cc-pV5Z basis was extended with a set of additional diffuse functions of each angular symmetry, according to the even-tempered scheme implemented in MOLPRO.~\cite{MOLPRO_brief}

\begin{figure}
\begin{center}
\includegraphics[scale=0.34]{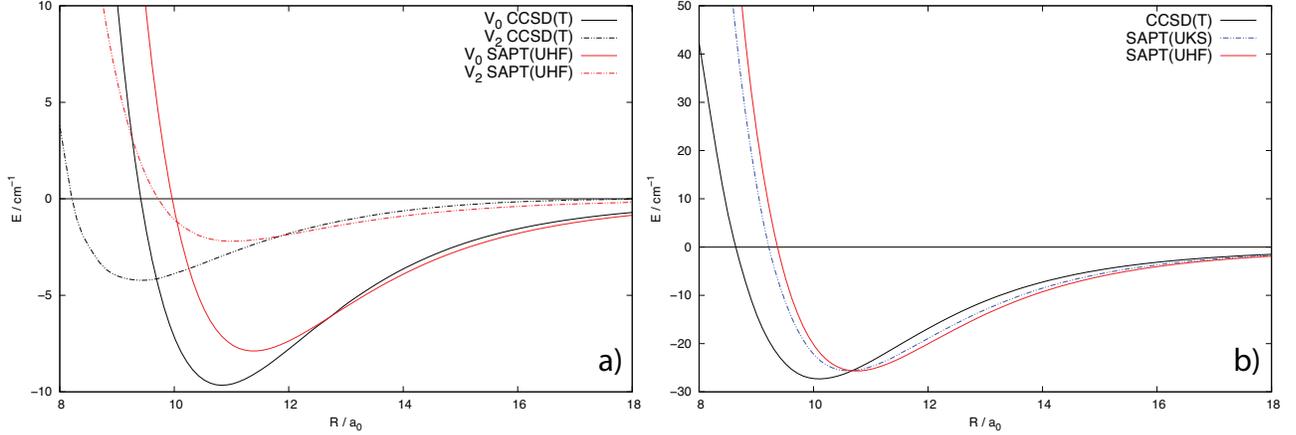}
\caption{a) Isotropic ($V_0$) and leading anisotropic terms ($V_2$)  of the He($^3$S)$\cdots$H$_2$ interaction, b) interaction potential for He($^3$S)$\cdots$Ar calculated at different levels of theory. See the text for description of the calculation details.}
\label{fig:v0heh2}
\end{center}
\end{figure}

\section{Numerical results}
\subsection{He($^{3}$S)$\cdots$X systems}
Fig.~\ref{fig:v0heh2} shows calculated potential energy curves for the He$^*$-X systems and the parameters of the interaction potentials, R$_e$, D$_e$ and C$_6$, are given in Table \ref{tab:he}. In the case of the He($^{3}$S)$\cdots$H$_2$ system (Fig.~\ref{fig:v0heh2}a) we have shown the isotropic ($V_0$) part of the interaction potential along with the leading anisotropic terms ($V_2$). 

We have investigated both SAPT(UHF) as well as SAPT(UKS) approaches as described in Ref.~\onlinecite{Hapka:2012}. Analysis of the $V_0$ potential parameters for the He($^{3}$S)$\cdots$H$_2$ system reveals that both SAPT schemes remain in a reasonably good agreement with UCCSD(T) reference values. The good performance of SAPT(UHF) can be attributed to the basis set optimization procedure used in Ref.~\onlinecite{Przybytek:2008}. Moreover, for a two-electron helium atom in its $^{3}$S state we found SAPT(UHF) to give a better description of electron spin density than SAPT(UKS). The erroneous behaviour of the PBE0 spin density is reflected, for instance, by the static polarizability. DFT calculations lead to the value of 258.79, whereas UHF predicts 317.04 a.u, much closer to the benchmark result of 315.63 a.u..~\cite{yan1998long}

UCCSD(T) places the minimum at 10.8 $a_0$ with the depth of 9.65 cm$^{-1}$, whereas SAPT(UHF) leads to well depth of 7.88 cm$^{-1}$ at larger distance of 11.4 $a_0$, which amounts to 18\% difference. These results are in good agreement with the potential derived from the experiment of Henson {\em et al.}~\cite{Henson:2012}, where the well depth of 8 cm$^{-1}$ has been found at $R_e$ of approximately 11.5 $a_0$. For comparison, the most recent semi-empirical potential from Refs.~\onlinecite{Martin:85, Martin:88} predicted the minimum at a similar distance, however its well depth twice as shallow (See Table~\ref{tab:he}).

The leading anisotropy of the potential energy surface, $V_2$ is very small. In the van der Waals minimum region both methods agree very well, however, the UCCSD(T) potential anisotropy is much larger in the short-range repulsive region. 

It is also interesting to compare results for the global minimum of the potential energy surface. Both methods give the minima at the linear geometry: UCCSD(T) predicts a well depth of 12.86 cm$^{-1}$ at 10.6 $a_0$, while SAPT leads to shallower minima. SAPT(UHF) gives a well depth of 9.28 cm$^{-1}$ at 11.4 $a_0$ and SAPT(UKS) gives minimum of 9.23 cm$^{-1}$ at 11.2 $a_0$. 

Similar to the He$^*\cdots$H$_2$ system, the well depth of the He$^*\cdots$Ar potential energy curve is consistent for all of the applied methods (within 7\%). The equilibrium distance of the SAPT potential curve is slightly larger than in case of CCSD(T) which is consistent with the behaviour of $V_0$ of the He($^{3}$S)$\cdots$H$_2$ potential.

In Tab. \ref{tab:he} we present the (isotropic) $C_6$ coefficients. Apart from comparing the calculated $C_6$ coefficients with reference values found in the literature, we have applied the Tang's formula~\cite{tang1969dynamic}:
\beq
C_{6,AB} = \frac{2\alpha_A^0 \alpha_B^0 C_{6,AA} C_{6,BB}}{(\alpha^0_A)^2 C_{6,BB} + (\alpha^0_B)^2 C_{6,AA}},
\eeq
where $\alpha^0_X$ stands for static polarizabilities and $C_{6,XY}$ denote $C_6$ coefficients for homo- and heteronuclear dimers. Static polarizabilities for He$^*$ ($\alpha_{He^*}^0$ = 315.63 a.u.) and H$_2$ ($\alpha_{H_2}^0$ = 5.18 a.u.) were taken from Refs.~\onlinecite{yan1998long,Korona:2006}, respectively. For argon atom we used $\alpha^0_{Ar}$ = 11.078 a.u. and $C_{6,ArAr}$ = 64.42 a.u. values from Ref.~\onlinecite{kumar2010dipole}. Finally, $C_{6,He^*He^*}$ = 3276.67 a.u. and $C_{6, H_2H_2}$ = 12.058 a.u. were found in Refs.~\onlinecite{zhang2006} and~\onlinecite{bishop93}, respectively.  
As summarized in Table \ref{tab:he} all of the applied methods lead to consistent results for $C_6$ coefficients and stay in agreement with previous theoretical prediction (see Refs.~\onlinecite{Victor:70} and~\onlinecite{Proctor:78}).

In order to test the He$^*\cdots$H$_2$ and He$^*\cdots$Ar potentials we calculated the position of the shape resonances in the elastic low-energy (up to 1 meV) scattering of He($^3$S) with the molecular hydrogen and argon.
Since the {\it ab initio} methods applied in this paper provide only the description of the entrance channel of PI process, we are not in a position to predict the reaction rates. Nevertheless, the enhancement of the PI reaction rates attributed by Henson {\em et al.}~\cite{Henson:2012} to shape resonances is entirely the {\em entrance-channel} effect, thus they correspond exactly to the calculated elastic scattering cross sections in the reactive scattering.
 
\begin{figure}
\begin{center}
\includegraphics[scale=0.68]{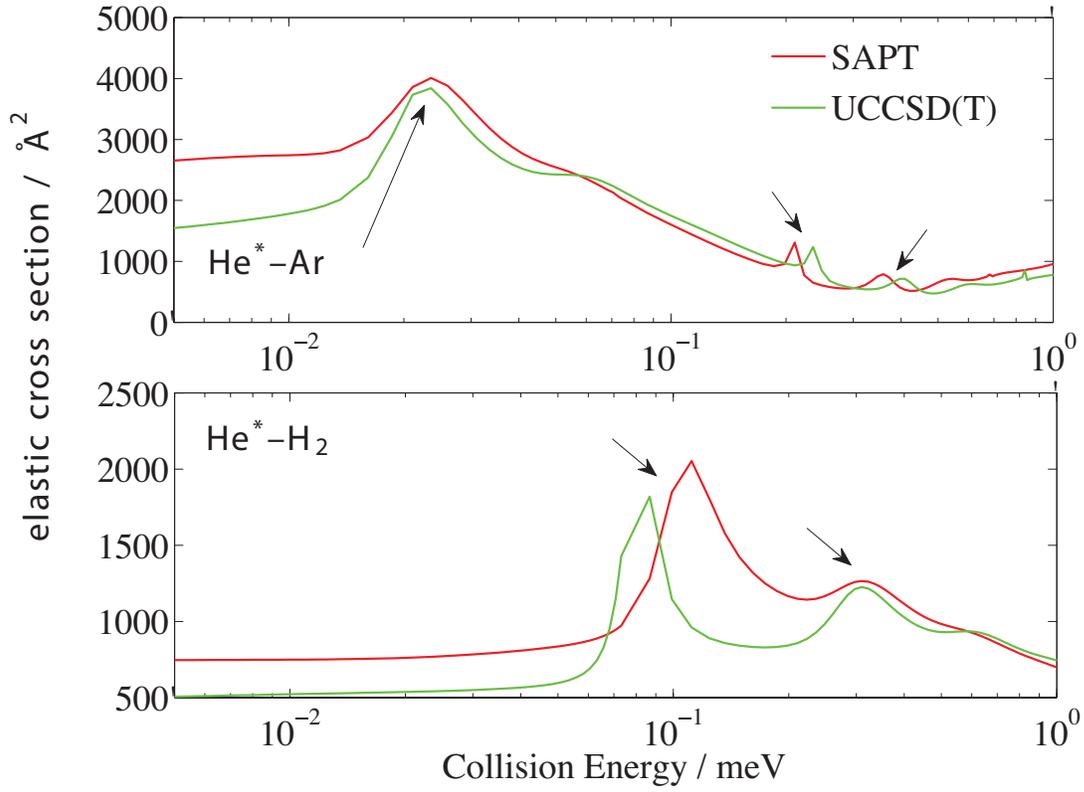}
\caption{Shape resonances in the elastic cross sections for the scattering of  He$^*\cdots$Ar (upper panel) and He$^*\cdots$H$_2$ (lower panel) systems. 
The arrows depict the strongest resonances which we compare to the experiment~\cite{Henson:2012}.}
\label{he:scat}
\end{center}
\end{figure}

Prediction of resonances is a particularly demanding test of the quality of the potential curve.
The calculated elastic cross sections for He$^*\cdots$H$_2$ and  He$^*\cdots$Ar systems are shown in the Fig.~\ref{he:scat}
and the overall pattern of the resonances is nearly identical for UCCSD(T) and SAPT potentials.
For the He$^*\cdots$H$_2$ system, for which we used only $V_0$ potential in scattering calculations,
we found strong shape-resonances: the first near 0.08 meV for UCCSD(T) and 0.1  meV for SAPT, and second at 0.3 meV for 
both potentials. They originate from the partial wave components with end-over-end quantum number $l=3$ and $4$, respectively.
In the experiment both resonances were found at lower energies, ca. 0.02 meV and 0.2 meV, respectively. 
Note that Henson {\em et al.}~\cite{Henson:2012} attributed those resonances to $l=5$ and $6$, respectively, whereas in our work they appear for $l=3$ and $4$. In contrast to what is shown by Henson {\em et al.}, our adiabatic curve associated with the $l=6$ partial wave exhibits no attractive minimum. 

In the case of He$^*\cdots$Ar system, both UCCSD(T) and SAPT potentials lead to slight shifts in resonance positions.
The position of the first shape resonance, at ca. 0.02 meV, is in a very good agreement with experimental findings.~\cite{Henson:2012} The second and third strong resonances in the experimental studies were found at approximately 0.1 meV and 0.3 meV, respectively. These features are, again, properly reproduced by calculations with the potentials obtained in this paper (peaks at ca. 0.2 meV and 0.4 meV).

\begin{threeparttable}[c]
\caption{Parameters of the isotropic part of the interaction potential, $V_0$, for He($^{3}$S)$\cdots$H$_2$ and He($^{3}$S)$\cdots$Ar interaction potential together with $C_6$ coefficients. TF denotes results obtained with Tang's formula}
\begin{tabular}{l p{4mm} c p{0mm} c c p{9mm} c p{0mm} c c}
\toprule
 & & \multicolumn{4}{c}{He($^{3}$S)$\cdots$H$_2$} && \multicolumn{4}{c}{He($^{3}$S)$\cdots$Ar} \\
Method     && R$_e$ [$a_0$]      && D$_e$ [cm$^{-1}$] & C$_{60}$ [a.u.]  && R$_e$ [$a_0$] && D$_e$ [cm$^{-1}$] & C$_6$ [a.u.] \\ \midrule
CCSD(T)    && 10.80  && 9.65   & 97.8                 && 10.09 && 27.31 & 214.1 \\
SAPT(UHF)  && 11.38 && 7.88   & 112.8                  && 10.75 && 25.68 & 220.0 \\
SAPT(UKS)  && 11.24  && 7.62  & 104.9                  && 10.61 && 25.63 & 206.8 \\  
TF.        && ---   && ---     & 100.2                  && ---   && ---    & 216.4 \\
Ref.       && 11.72\tnote{a)}  && 4.72\tnote{a)} & 109.6\tnote{b)} && 9.77\tnote{c)}  && 41.27\tnote{c)}  & 220.9\tnote{d)} \\ \bottomrule
\end{tabular}
 \begin{tablenotes}
\item [a)] See Ref.~\onlinecite{Martin:85, Martin:88}
\item [b)] See Ref.~\onlinecite{Victor:70} 
\item [c)] See Ref.~\onlinecite{Siska:79}
 \item [d)] See Ref.~\onlinecite{Proctor:78}
 \end{tablenotes}
\label{tab:he}
 \end{threeparttable}

\subsection{Ne($^3$P)$\cdots$X systems}
For the Ne($^3$P)$\cdots$Ar complex we were not able to obtain convergent CCSD(T) results and SAPT seems to be the only method allowing for an insight into the nature of Ne($^3$P)$\cdots$Ar interaction. In order to test the performance of SAPT in this case, we decided to choose a similar Ne($^3$P)$\cdots$He system, in which PI does not occur.
For both of the studied complexes the SAPT(UKS) results proved to be of practically the same accuracy as the SAPT(UHF) ones,
hence, only SAPT(UHF) potentials are discussed.

\begin{figure}
\begin{center}
\includegraphics[scale=0.34]{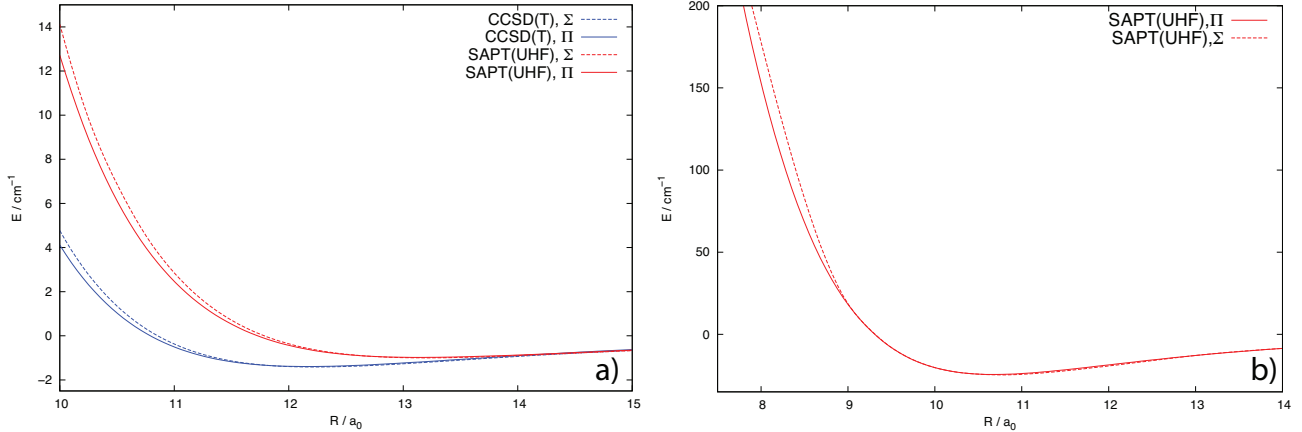}
\caption{Potential energy curves for  a) Ne($^3$P)$\cdots$He, b) Ne($^3$P)$\cdots$Ar calculated at different levels of theory. See the text for description of basis sets.}
\label{fig:nehew1}
\end{center}
\end{figure}

In Fig. \ref{fig:nehew1}a and Table \ref{tab:ne} we compare UCCSD(T) and SAPT(UHF) interaction potentials for $^3\Sigma$ and $^3\Pi$ states of Ne($^3$P)$\cdots$He.
The interaction potentials for that system are extremely weak: the well depths are very shallow, on the order of 1 cm$^{-1}$, and the minima occur at a very long range ($>12$ $a_0$).
The performance of SAPT(UHF) in comparison with UCCSD(T) is similar to the He*$\cdots$H$_2$ case. 
The values of the potentials are very small due to strong cancellation of exchange and dispersion energies -- thus, they are strongly susceptible to relative errors. SAPT(UHF) correctly recovers the $\Sigma-\Pi$ anisotropy in this system, which is around 0.03 cm$^{-1}$ in the minimum and therefore should be regarded as a subtle effect.
 
Fig. \ref{fig:nehew1}b and Table \ref{tab:ne} present results for Ne($^3$P)$\cdots$Ar. SAPT predicts a minimum at 10.6 bohr for $^3\Sigma$ and 10.8 bohr for $^3\Pi$ state, 24.68 cm$^{-1}$ and 24.40 cm$^{-1}$ deep, respectively.
Such extremely weak anisotropy of Ne$^*$ is, in fact, not suprising. A similar effect was observed for lanthanides and some transition metals in their ground state where it was attributed to the effective screening of the doubly-filled outermost $s$ shell, see, e.g., Refs.~\onlinecite{rajchel:2007, buchachenko2006ab, buchachenko2006van}. In the metastable neon atom such screening is even more pronounced. A weak anisotropy of Ne($^3P$) can be visualized in the following orbital picture: the electron lying in the outer 3$s$ orbital effectively screens the inner 2$p$ shell, which has much lower $\braket{r}$ value, causing substantial suppression of the anisotropy.
It also manifests itself in a small quadrupole moment of the metastable Ne atom (0.092 a.u. calculated as expectation value with the CISD method),
an order of magnitude smaller than quadrupole moments of the other second-row $P$-state atoms (e.g. 0.92 for O $^3$P atom).
In their work Gregor and Siska~\cite{Gregor:81} also postulated that effect and therefore assumed a single triply degenerate $\Sigma$--$\Pi$ 
component in their model potential for Ne($^3$P)$\cdots$Ar. They placed the minimum at 9.45 $a_0$ with the well depth of 43.72 cm$^{-1}$.
The $C_6 = 225.11$ a.u. obtained here with SAPT(UHF) agrees fairly well with the value of $204.68$ a.u. calculated using the theory of Proctor and Stwalley~\cite{Proctor:78} and given by Gregor and Siska.~\cite{Gregor:81}

\begin{table}
\caption{Global minima for Ne($^{3}$P)$\cdots$He/Ar predicted at different levels of theory.}
\begin{tabular}{l p{4mm} c p{0mm} c c p{9mm} c p{0mm} c c}
\toprule
 & & \multicolumn{4}{c}{Ne$\cdots$He ($^2\Sigma$)} && \multicolumn{4}{c}{Ne$\cdots$Ar ($^2\Sigma$)} \\
Method     && R$_e$ [$a_0$]      && D$_e$ [cm$^{-1}$] & C$_{6}$ [a.u.]  && R$_e$ [$a_0$] && D$_e$ [cm$^{-1}$] & C$_6$ [a.u.] \\ \midrule
CCSD(T)    && 12.22 && 1.414     & 24.6 && ---  && ---   & --- \\
SAPT(UHF)  && 13.25 && 0.998     & 29.1 && 10.8 && 24.68 & 231.24 \\ \midrule
 & & \multicolumn{4}{c}{Ne$\cdots$He ($^2\Pi$)} && \multicolumn{4}{c}{Ne$\cdots$Ar ($^2\Pi$)} \\
CCSD(T)    && 12.15 && 1.393 & 24.3 && ---  && ---   & --- \\
SAPT(UHF)  && 13.14 && 0.980 & 28.3 && 10.6 && 24.40 & 225.11 \\ \bottomrule
\end{tabular}
\label{tab:ne}
\end{table}

\section{Conclusions}
We have investigated the potential energy curves of the He$^*\cdots$H$_2$ ($V_0$ and $V_2$) and He$^*\cdots$Ar systems with two different methods: UCCSD(T) and SAPT(UHF/UKS). In both cases we have obtained a consistent picture of the interaction with minor differences in predicted $D_e$ and $R_e$. We have used these potentials to calculate the positions of the shape resonances in low-energy scattering obtaining a good agreement with recently reported state-of-the-art scattering experiment of Henson {\em et al.}.~\cite{Henson:2012}

We have also investigated two systems containing metastable Ne atoms: in case of Ne$^*\cdots$He system we have obtained a good agreement between UCCSD(T) and SAPT which suggests that the SAPT description of Ne$^*$Ar dimer should be correct. We have revealed strikingly small electronic orbital angular momentum anisotropy of these potentials which can be explained by a strong screening of the open $2p$ shell by the outermost, singly occupied $3s$ shell.

As far as the performance of SAPT is concerned, we have exploited the unique feature of this method: fixing the number of electrons at each monomer to prevent an electron-hopping. This approach is expected to be general, provided that proper reference states of the monomers can be obtained (either Hartree-Fock or Kohn-Sham wave functions). Another prerequisite of applying SAPT must be its convergent behaviour in the lowest orders. It cannot be taken for granted in general, for instance when both monomers are largely polarizable -- in this last case, one could apply appropriately modified Pauli-Blockade method~\cite{Rajchel:2010} combined with second-order dispersion contribution obtained with SAPT.

In the near future we hope to take advantage of SAPT to survey the potential energy surfaces of He$^*$-paramagnetic molecule (or atom) systems exploring the possibility of their sympathetic cooling.

\section{Acknowledgments}
M. H. was supported by ``Towards Advanced Functional Materials and Novel Devices - Joint UW and WUT International PhD Programme'' Project operated within the Foundation for Polish Science MPD Programme, implemented as a part of the Innovative Economy Operational Programme (EU European Regional Development Fund).
P. S. Z. acknowledges funding from the Homing Plus programme (Project No. 2011-3/14) of the Foundation for Polish Science, which is co-financed by the European Regional Development Fund of the European Union.
J. K. would like to thank for the financial support through the United States National Science Foundation grant No: CHE-1213332 to Prof. Millard Alexander.
G. Ch. is grateful to the Polish Ministry of Science and Higher Education, Grant No. N204 248440, and by the National Science Foundation (US), Grant No. CHE-1152474.

\end{document}